\documentclass[preprint]{aastex631}

\usepackage{amsmath}
\usepackage{gensymb}

\begin{document}
\title{A new shock in the pre-merging cluster pair 1E2215-2216}

\author[0009-0001-4545-9603]{Yanling Chen}
\affiliation{Department of Astronomy, Tsinghua University, Beijing 100084, China}

\author{Liyi Gu}
\affiliation{SRON Netherlands Institute for Space Research, Niels Bohrweg 4, 2333 CA Leiden, The Netherlands}

\author{Aurora Simionescu}
\affiliation{SRON Netherlands Institute for Space Research, Niels Bohrweg 4, 2333 CA Leiden, The Netherlands}

\author{Chunyang Jiang}
\affiliation{Department of Astronomy, Tsinghua University, Beijing 100084, China}

\author{Rui Huang}
\affiliation{Department of Astronomy, Tsinghua University, Beijing 100084, China}
\affiliation{Department of Astronomy, University of Michigan, Ann Arbor, MI, 48109-1107, USA}

\author{Wei Cui}
\affiliation{Department of Astronomy, Tsinghua University, Beijing 100084, China}

\begin{abstract}

The galaxy cluster pair 1E2216.0-0401 and 1E2215.7-0404 represents a major cluster merger in its early stages, a phase that has been scarcely explored in previous studies. Within this system, both axial and equatorial merger shocks have been identified. Recent \textsl{XMM-Newton} observations of the southern region of the cluster pair have increased the total exposure time to approximately 300\,ks, enhancing the sensitivity to detect faint shock features in the cluster outskirts. Through a combined analysis of \textsl{XMM-Newton} and \textsl{Chandra} data, including both imaging and spectral techniques, a new shock front has been identified at approximately 2.3 arcminutes south of the X-ray brightness peak of 1E2215. This shock front exhibits a surface brightness ratio of \( 1.33 \pm 0.07 \) and a temperature ratio of \( 1.22^{+0.13}_{-0.14} \) in \textsl{XMM-Newton}, consistent with \textsl{Chandra} results. The Mach number, independently calculated from both the temperature and surface brightness discontinuities, yields consistent values of \( \mathcal{M} \approx 1.2 \). The age, velocity, and spatial distribution of this shock suggest that it shares a common physical origin with the previously identified equatorial shock.

\end{abstract}


\section{Introduction}

Galaxy cluster mergers are the most energetic events in large-scale structure formation \citep{Kravtsov12}, releasing kinetic energy on the order of $10^{64}$ ergs, which dissipates over a timescale of approximately 1\,Gyr \citep{Botteon18}. A portion of this energy is converted into thermal energy of the intracluster medium (ICM) through collision-induced shocks (\citealt{Markevitch07}), while the remainder is channeled into non-thermal processes, including cosmic ray acceleration, ICM turbulence generation, and magnetic field amplification (\citealt{Weeren19, Feretti12}). However, the precise distribution of dynamical energy among these processes remains uncertain, which is closely tied to the cosmologically significant challenge of accurately determining the total masses of galaxy clusters (\citealt{Loeb94}). As shocks propagate supersonically into the cluster outskirts, they induce discontinuities in the ICM \citep{Molnar16}, which are observed as surface brightness edges and temperature jumps in X-ray data. Using the Rankine-Hugoniot conditions (\citealt{Landau59}), the Mach number of the shocks can be estimated from these X-ray features, providing insights into their speed and strength \citep{Markevitch07}. By combining shock velocities and temperature jumps, the kinetic energy of the shocks can be derived and compared with the thermal energy, enabling a detailed analysis of energy conversion processes.

The 1E2215-2216 system is in an early pre-merger phase, located at a redshift of \( z \sim 0.094 \). The X-ray brightness peaks of the two clusters are located at (22:18:39.54, -3:46:44.47) for 1E2216 and (22:18:14.98, -3:49:36.92) for 1E2215, with a projected separation of 640\,kpc (approximately 7.2$'$). Based on the mass-temperature relation, their total masses are estimated to be approximately \( 2.5 \times 10^{14}~M_{\odot} \) and \( 3.2 \times 10^{14}~M_{\odot} \), respectively \citep{Gu19}. Previous studies identified an equatorial shock at approximately 4.2$'$ from the merger axis, with a Mach number of \( \mathcal{M} = 1.6 \pm 0.3 \) \citep{Gu19}, marking the first observation of an equatorial shock in an early-stage merging cluster system. Additionally, using \textsl{Suzaku} data, an axial shock was detected at approximately 4 arcminutes from 1E2216 along the merger axis, with a Mach number of \( \mathcal{M} = 1.4 \pm 0.1 \) \citep{Akamatsu16}. Diffuse radio emission was also observed between the two clusters, likely originating from fossil electrons accerlerated by the axial shock.

In this study, we utilized the new and archival \textsl{XMM-Newton} data along with archival \textsl{Chandra} to further investigate the gas properties of the pre-merger cluster pair 1E2216 and 1E2215 through imaging and spectral analysis. The combined exposure time of the new and archival \textsl{XMM-Newton} data in the southern region of the cluster pair reaches approximately 300\,ks, enabling new discoveries. The cosmological model adopted in this study is the standard $\Lambda$CDM model, with parameters \( H_0 = 70 \, \rm{km\, s^{-1}\, Mpc^{-1}} \), \( \Omega_{M} = 0.3 \), and \( \Omega_{\Lambda} = 0.7 \). Under this cosmology, an angular scale of 1$’$ corresponds to 104.8\,kpc at the redshift of 0.094, where the 1E2215/2216 system is located. Here, \( r_{200c} \) and \( r_{500c} \) represent the radii within which the average density is approximately 200 and 500 times the critical density of the universe, respectively. Unless otherwise specified, all reported uncertainties are at the \( 1\sigma \) confidence level.

\section{Observations}

\subsection{XMM-Newton}

\begin{figure}[ht!]
\plotone{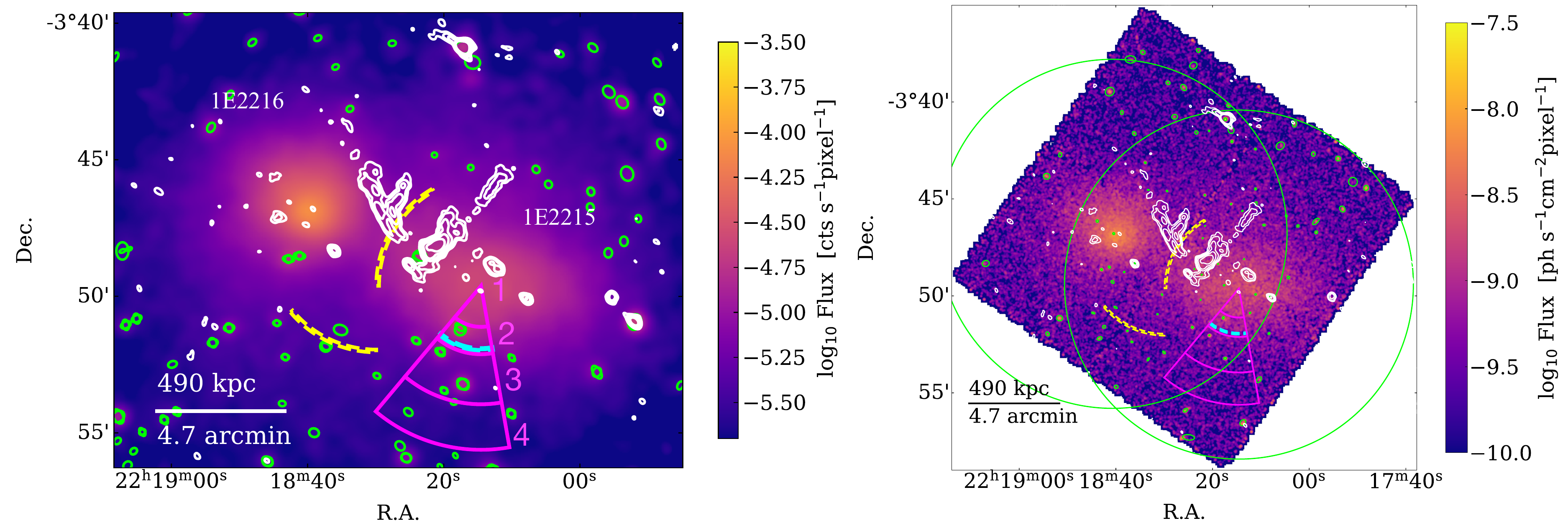}
\caption{Left panel‌: Zoom-in \textsl{XMM-Newton} image in the [0.5-2.0]\,keV band (extracted from the red box region in Fig.\ref{fig:xmm-overall}), with white contours indicating 235 MHz radio intensity observed by GMRT. Cyan dashed curves mark surface brightness edges identified through surface brightness profile analysis in this work, yellow dashed curves denote edges from previous studies, green ellipses outline point source exclusion regions, and magenta sectors correspond to spectral extraction and image analysis regions in the \textsl{XMM-Newton} data. The spectral extraction sectors centered on 1E2215 from inner to outer are numbered as 1, 2, 3, 4. Right panel‌: Adaptively smoothed \textsl{Chandra} image in the same energy band, with particle background subtracted and vignetting corrected. All annotations are the same as in the left panel; in addition, the large green circles mark the inner limits of the background region.
\label{fig:xmm-zoomin}}
\end{figure}

The \textsl{XMM-Newton} observations used for the 1E2215-2216 system are summarized in table\,\ref{tab:obs_xmm}. Among these observations, 0800380101 targets the center of the cluster pair, while observations 0881210101/201, newly acquired in 2021, were offset to the south of the system. Data reduction primarily followed the Extended Science Analysis System (ESAS) pipeline\footnote{\url{https://pages.jh.edu/kkuntz1/xmm-esas.pdf}} \citep{Snowden08}, using the Scientific Analysis System (XMMSAS) v21.0.0 for processing. However, due to unresolved issues with PN spectra extraction in SAS v21.0.0, SAS v20.0.0 was employed for PN spectra extraction \citep{Rossetti24}. The tasks \texttt{emproc} and \texttt{epproc} were used to apply the latest calibration files released in January 2021, and out-of-time (OOT) event files for PN were generated using \texttt{epproc}. Events with patterns greater than 4 for MOS and greater than 12 for PN were excluded. Anomalous chips in MOS, marked with status 'O', 'U', or 'B', were excluded using the \texttt{emanom} task. Good time intervals were selected using \texttt{espfilt} from the light curve. We calculated the IN/OUT field-of-view (FOV) count rate ratio for MOS2 to assess the quiescent soft proton contribution during the observations \citep{Deluca04}, which indicates that the data were largely free from quiescent soft proton contamination. Therefore, the soft proton contribution was ignored in the imaging analysis. Point sources were detected using the task \texttt{edetect\_stack}, combining all observations and detectors, and were visually inspected. Only point sources with likelihood larger than 6 are excluded. The \texttt{region} command was used to generate exclusion regions around the point sources with background fraction 0.1 in contour mode. The excluded point source regions are marked with green ellipses in Fig.\ref{fig:xmm-zoomin} (zoom-in view) and Fig.\ref{fig:xmm-overall} (overall view). Spectra and images were extracted using the tasks \texttt{mos}/\texttt{pnspectra}. Particle background images and spectra were generated using \texttt{mosback} and \texttt{pn\_back}, based on Filter Wheel Closed (FWC) data. All images from the three detectors (MOS1, MOS2, and PN) were combined by \texttt{comb} task and co-added by \texttt{emosaic}. Finally, the background-subtracted data images were divided by the exposure maps to generate a vignetting-corrected count rate map, as shown in Fig.\ref{fig:xmm-zoomin}.

\begin{deluxetable*}{ccccccc}
\label{tab:obs_xmm}  
\tablecaption{\textsl{XMM-Newton} observations and clean exposure time used in the paper}    
\tablehead{\colhead{Telescope} & \colhead{OBSID} & \colhead{Total} & \colhead{MOS1} & \colhead{MOS2} & \colhead{PN} & \colhead{MOS2 IN/OUT ratio} }
\startdata
& 0800380101 & 139.0 & 80.1 & 82.1 & 67.5 & 1.08 \\  
\textsl{XMM-Newton}& 0881210101 & 79.6  & 70.5 & 71.4 & 63.1 & 1.02 \\
& 0881210201 & 81.9  & 36.8 & 39.3 & 29.1 & 1.03 \\
\textsl{Chandra} & 20778,21131/2/3/4 & 121.6 & & & &  \\
\enddata 
\end{deluxetable*}

\subsection{Chandra}

The \textsl{Chandra} observations used for the 1E2215-2216 system including observation IDs 20778 and 21131/2/3/4, with a total filtered exposure time of 121.6\,ks, and individual exposures of 31.3\,ks, 24.7\,ks, 31.5\,ks, 19.1\,ks, and 15.0\,ks for each observation, respectively. Data reduction was performed using the \textsl{Chandra} Interactive Analysis of Observations (CIAO) v4.15 and the calibration database (CALDB) 4.11.2. Event and auxiliary files were generated using the \texttt{chandra\_repro} task\footnote{\url{https://cxc.cfa.harvard.edu/ciao/threads/acisbackground/}}, with default settings except for setting `check\_vf\_pha=yes'. Particle background images and spectra were generated using stowed event files, and the \texttt{lc\_clean} task was applied to filter the light curve of stowed events in the [0.5-7.0]\,keV band, excluding time intervals with count rates exceeding 3 sigma. The images were reprojected and merged, as shown in right panel of Fig.\ref{fig:xmm-zoomin}. Point source detection was performed via \texttt{wavdetect} with scale parameters set to 2, 4, 8, 16, 24, 32 and 48, while the exposure map was incorporated to mitigate false detections. The source catalog generated by \texttt{wavdetect} was then fed into the \texttt{roi} task to create exclusion regions with a background factor of 0.1 (green ellipses), which are excluded from further analysis. To eliminate CCD gaps, adaptive smoothing using \texttt{dmimgadapt} was applied to the exposure map, particle background map, and source map. The region beyond $\sim 1.1\rm r_{200c}$ (9 arcmin; see green circles) of each cluster center is defined as the background region to constrain the LHB, GH and CXB in the spectral analysis. The spectra are extracted from the \textsl{Chandra} with the same regions in \textsl{XMM-Newton} using specextract.


\section{Results}

\subsection{Surface brightness profile}

\begin{figure}[ht!]
\plotone{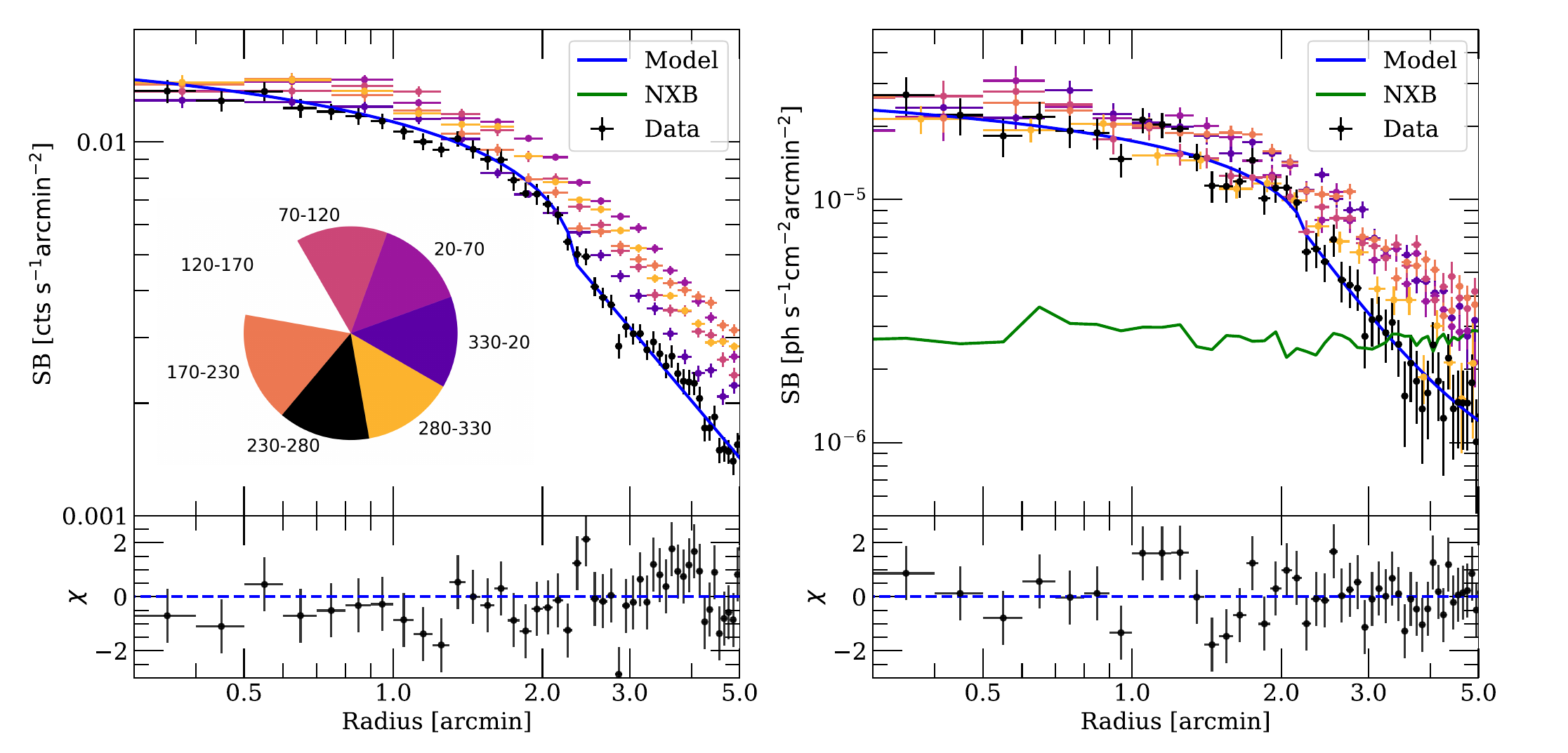}
\caption{The surface brightness radial profiles were extracted from azimuthal sectors centered on the 1E2215 brightness peak using \textsl{XMM-Newton} data (left panel) and \textsl{Chandra} data (right panel), with all profiles extracted in the [0.5-2.0]\,keV energy band. Black data points represent profiles from the 230$\degree$-280$\degree$ sector, which is marked by a magenta sector in figure~\ref{fig:xmm-zoomin}. The inset in the left panel shows the opening angle of the sectors, with sector colors matching the data point colors, where 0\textdegree corresponds to the west. Colored data points indicate profiles from 170$\degree$-230$\degree$ (orange), 70$\degree$-120$\degree$ (yellow), 20$\degree$-70$\degree$ (cyan), 330$\degree$-20$\degree$ (purple), and 280$\degree$-330$\degree$ sectors (see color legend in left panel). The 120$\degree$-170$\degree$ sector is excluded due to its alignment toward another galaxy cluster. Blue curves show the best-fit brightness distribution models, while green curves display radial profiles extracted from particle background images. The particle background level for \textsl{XMM-Newton} is below $10^{-3} \rm~ cts\ s^{-1}arcmin^{-2}$, rendering it invisible in the plot.
\label{fig:sbp-prof}}
\end{figure}

In Fig.\ref{fig:sbp-prof}, surface brightness profiles are extracted in the [0.5-2.0]\,keV energy band from sectors centered on the brightness peak of 1E2215. The 0$\degree$ direction corresponds to the west.  Five sectors with opening angles of 170$\degree$-230$\degree$, 70$\degree$-120$\degree$, 20$\degree$-70$\degree$, 330$\degree$-20$\degree$, and 280-330$\degree$ are plotted in distinct colors, contrasting with the reference 230$\degree$-280° sector (black), which is the sector where the new shock front is detected at $\sim 2.3’$ from 1E2215's center (marked in magenta in Fig.\ref{fig:xmm-zoomin}). The 120$\degree$-170$\degree$ sector is excluded due to alignment with galaxy cluster 1E2216. The 230$\degree$-280$\degree$ profile exhibits a steeper brightness decline than other sectors, a feature also present in the \textsl{Chandra} image of the same region.

We fitted this edge using a projected broken power law model with \texttt{pyproffit} \citep{Eckert20} \footnote{\url{https://pyproffit.readthedocs.io/en/latest/_modules/pyproffit}}, defined as 

\begin{equation}
I\left(r\right) = I_0 \int F(\omega)^2 dl +B
\end{equation}
\begin{equation}
F(\omega) = 
\begin{cases} 
\omega^{-\alpha_{\text{in}}}, & \text{if } \omega < r_{\text{break}}, \\ 
\frac{n_{\text{out}}}{n_{\text{in}}} \omega^{-\alpha_{\text{out}}}, & \text{otherwise}.
\end{cases}
\end{equation}

Here, $I\left(r\right)$ represents the surface brightness intensity as a function of the projected distance $r$ from the center. The 3D true distance $\omega$ is defined as $\omega = \sqrt{r^2 + l^2}$, where $l$ denotes the line of sight distance. $I_0$  corresponds to the surface brightness normalization factor, $B$ represents the background level of the model, and the power-law indices \( \alpha_{\rm in} \) and \( \alpha_{\rm out} \) characterize the density distribution interior and exterior to the edge, respectively. The parameter \( r_{\rm break} \)  marks the projected radius of the potential edge, and, and \( n_{\rm in}/n_{\rm out} \) quantifies the density jump across the edge (i.e., the ratio of inner-to-outer densities). This broken power-law model is projected along the line of sight, convolved with the point spread function (PSF), and applied to both telescope datasets using linear binning with an angular size of 10$''$ applied to both telescopes.

\begin{deluxetable*}{cccccccc}
\label{tab:sbp-para}  
\tablecaption{Fitting parameter of the surface brightness edges in [0.5-2.0]\,keV band}    
\tablehead{\colhead{Instrument} & \colhead{$r_{\rm break}$ [arcmin]} & \colhead{$\alpha_1$} & \colhead{$\alpha_2$} & \colhead{$n_{\rm in}/n_{\rm out}$} & \colhead{B} & $I_0$ & \colhead{$\chi^2$ (d.o.f)} } 
\startdata
XMM-Newton & $2.32 \pm 0.08$ & $0.28 \pm 0.08$ & $1.29 \pm 0.04$ & $1.33 \pm 0.07$ & $10^{-2.70 \pm 0.05}$ & $10^{-2.50 \pm 0.05}$ & 16 (17)\\
Chandra    & $2.25 \pm 0.01$ & $0.29 \pm 0.12$ & $1.97 \pm 0.25$ & $1.19 \pm 0.13$ & $10^{-5.99 \pm 0.05}$ & $10^{-5.28 \pm 0.05}$ & 29 (26)\\
\enddata 
\end{deluxetable*}

The best-fit model confirms the surface brightness jump at the southern edge, and the results are summarized in Tab.\ref{tab:sbp-para} and in Fig.\ref{fig:sbp-prof}. The measured brightness jumps for \textsl{XMM-Newton} ($1.33 \pm 0.07$) and \textsl{Chandra} ($1.19 \pm 0.13$) agree within 1$\sigma$. For both telescopes, the broken power-law model provides a better fit than the beta model. For \textsl{XMM-Newton}, the fitting statistics $\chi^2$ for the broken power-law model and the beta model are 16.4 (17 d.o.f) and 33.0 (19 d.o.f), respectively. For \textsl{Chandra}, the fitting statistics $\chi^2$ for the broken power-law model and the beta model are 29.2 (26 d.o.f) and 32.7 (28 d.o.f), respectively. We examined the best fitted model with penalty of the model parameter numbers using the Bayesian Information Criterion (BIC, \citealt{kass1995}). BIC indicates a strong evidence that the broken power-law model is the better fit in the \textsl{XMM-Newton} ($\rm BIC_{bkn}-BIC_{beta} = -6.23$) data of the 230$^{\circ}$-280$^{\circ}$ sector. However, \textsl{Chandra} shows that a beta model is slightly favored compared to the broken power law in this sector ($\rm BIC_{bkn}-BIC_{beta} = 1.64$), which might be due to the low SNR (signal-to-noise ratio) of the \textsl{Chandra} data. The $\Delta \rm BIC$ for all surface brightness profiles are listed in Tab.\ref{tab:bic} in the Appendix.

\subsection{Spectral properties}
We extracted spectra from sectors with an opening angle of 230$\degree$-280$\degree$ centered on 1E2215 in the \textsl{XMM-Newton} and \textsl{Chandra} data to study the thermodynamic properties of the plasma across the brightness jump (magenta sectors in Fig.\ref{fig:xmm-zoomin}). The radii of the sectors were chosen such that the SNR for each region exceeds 50, with radii of 1.5$'$, 2.5$'$, 4.3$'$, and 6.0$'$, numbered as sector 1, 2, 3, 4. For each sector, we used the \texttt{mos}/\texttt{pnspectra} tasks to extract the redistribution matrix files (RMFs), ancillary response files (ARFs), and the spectra themselves. The spectra were optimally binned using the method from \cite{kaastra16}, and the data were fitted using the C-statistic estimator \citep{humphrey09}.

We performed the spectral analysis using Xspec v12.13.0, adopting a composite model that includes particle background, source components, and sky background components. The sky background modeling is detailed in Appendix \ref{app:spec-mdls}. For MOS, the fitting was performed over the energy range [0.5-12.0]\,keV, while for PN, it was done over the range [0.7-14.0]\,keV. The [7.0-9.2]\,keV band in the PN spectrum contains Cu and Zn instrumental lines and was excluded during the fitting process. We used Gaussian models to fit the Al K$\alpha$ line (1.486\,keV) in both the MOS and PN spectra, as well as the Si K$\alpha$ line (1.739\,keV) in the MOS spectrum. The elemental abundances were referenced to the solar abundance table from \citealt{asplund09}. The normalization of the sky background was adjusted based on the ratio of the source region area to the background region area. The particle background model was smoothed and included as an additive component, with the normalization adjusted based on the ratio of the count rates in the [10.0-12.0]\,keV band of the source and background data. For the PN spectra, the OOT event spectrum from the PN instrument was smoothed using a similar method and used as a model in the spectral analysis. The source gas was modeled using the APEC model. We averaged the temperature results for three normalization levels of the cosmic X-ray background (CXB) and its errors accordingly. When calculating the gas density of the galaxy cluster, we assumed that the plasma in the sector region had a cylindrical shape, with a depth equal to the cluster's diameter (approximately 2\,Mpc), and used the normalization of the APEC model from the source fit to calculate the spatially averaged density based on this morphological assumption. The normalization here is defined according to the APEC model in Xspec, i.e., $\frac{10^{-14}}{4\pi\left[D_{A}\left(1+z\right)\right]^2\int n_e n_H dV}$, with $n_H = 1.2 n_e$. For \textsl{Chandra}, we fitted the data over the energy range of [0.7-10.0]\,keV, and all the model settings are the same as \textsl{XMM-Newton} except for the soft proton component, intrumental lines, and OOT events. 

\begin{deluxetable*}{ccccccc}
\label{tab:spec_props}
\tablecaption{ The best-fit kT and density in \textsl{XMM-Newton}}    
\tablehead{\colhead{CXB-set} & \colhead{Property} & \colhead{1} & \colhead{2} & \colhead{3} & \colhead{4} } 
    
\startdata
CXB & kT [keV]  & $6.58_{-0.36}^{+0.50}$ & $6.47_{-0.39}^{+0.56}$ & $5.30_{-0.34}^{+0.48}$& $5.07_{-0.49}^{+0.91}$ \\
       & $n_e$ [10$^{-4}$ cm$^{-3}$]  & $6.92_{-0.82}^{+0.86}$ & $4.55_{-0.57}^{+0.59}$ & $2.51_{-0.33}^{+0.33}$ & $1.86_{-0.31}^{+0.31}$  \\
CXB-$1\sigma$ & kT [keV]  & $6.66_{-0.42}^{+0.48}$ & $6.54_{-0.50}^{+0.53}$ & $5.47_{-0.36}^{+0.32}$& $5.35_{-0.49}^{+0.97}$ \\
       & $n_e$ [10$^{-4}$ cm$^{-3}$]  & $6.95_{-0.84}^{+0.83}$ & $4.56_{-0.57}^{+0.60}$ & $2.52_{-0.33}^{+0.33}$ & $1.88_{-0.31}^{+0.31}$  \\
CXB+$1\sigma$ & kT [keV]  & $6.51_{-0.37}^{+0.52}$ & $6.43_{-0.40}^{+0.56}$ & $5.24_{-0.33}^{+0.49}$& $4.97_{-0.49}^{+0.91}$ \\
       & $n_e$ [10$^{-4}$ cm$^{-3}$]  & $6.08_{-0.74}^{+0.76}$ & $4.53_{-0.58}^{+0.59}$ & $2.49_{-0.33}^{+0.34}$ & $1.82_{-0.31}^{+0.30}$  \\
\enddata 
\end{deluxetable*}

\begin{deluxetable*}{ccccccc}
\label{tab:spec_props_chandra}
\tablecaption{The best-fit kT and density in \textsl{Chandra}}    
\tablehead{\colhead{CXB-set} & \colhead{Property} & \colhead{1} & \colhead{2} & \colhead{3} } 
    
\startdata
CXB & kT [keV]  &
$7.88_{-0.79}^{+2.24}$ & 
$7.59_{-1.01}^{+2.94}$ &
$4.94_{-0.64}^{+1.84}$ \\
& $n_e$ [10$^{-4}$ cm$^{-3}$]  & 
$5.91_{-1.05}^{+1.07}$ & 
$4.12_{-0.77}^{+0.83}$ & 
$2.24_{-0.56}^{+0.51}$  \\
CXB-$1\sigma$ & kT [keV]  & 
$7.90_{-0.80}^{+2.28}$ & 
$7.63_{-1.03}^{+2.72}$ & 
$5.05_{-0.68}^{+1.71}$ \\
& $n_e$ [10$^{-4}$ cm$^{-3}$]  & 
$5.92_{-1.04}^{+1.09}$ & 
$4.13_{-0.77}^{+0.84}$ & 
$2.27_{-0.56}^{+0.53}$  \\
CXB+$1\sigma$ & kT [keV]  & 
$7.86_{-0.77}^{+2.29}$ & 
$7.55_{-1.02}^{+2.90}$ & 
$4.86_{-0.67}^{+1.74}$ \\
& $n_e$ [10$^{-4}$ cm$^{-3}$]  & 
$5.90_{-1.05}^{+1.07}$ & 
$4.10_{-0.77}^{+0.84}$ & 
$2.22_{-0.56}^{+0.52}$  \\
\enddata 
\tablecomments{ The kT of region 4 is not constrained by \textsl{Chandra} data, and therefore is not listed.}
\end{deluxetable*}

\begin{figure}[ht!]
\plotone{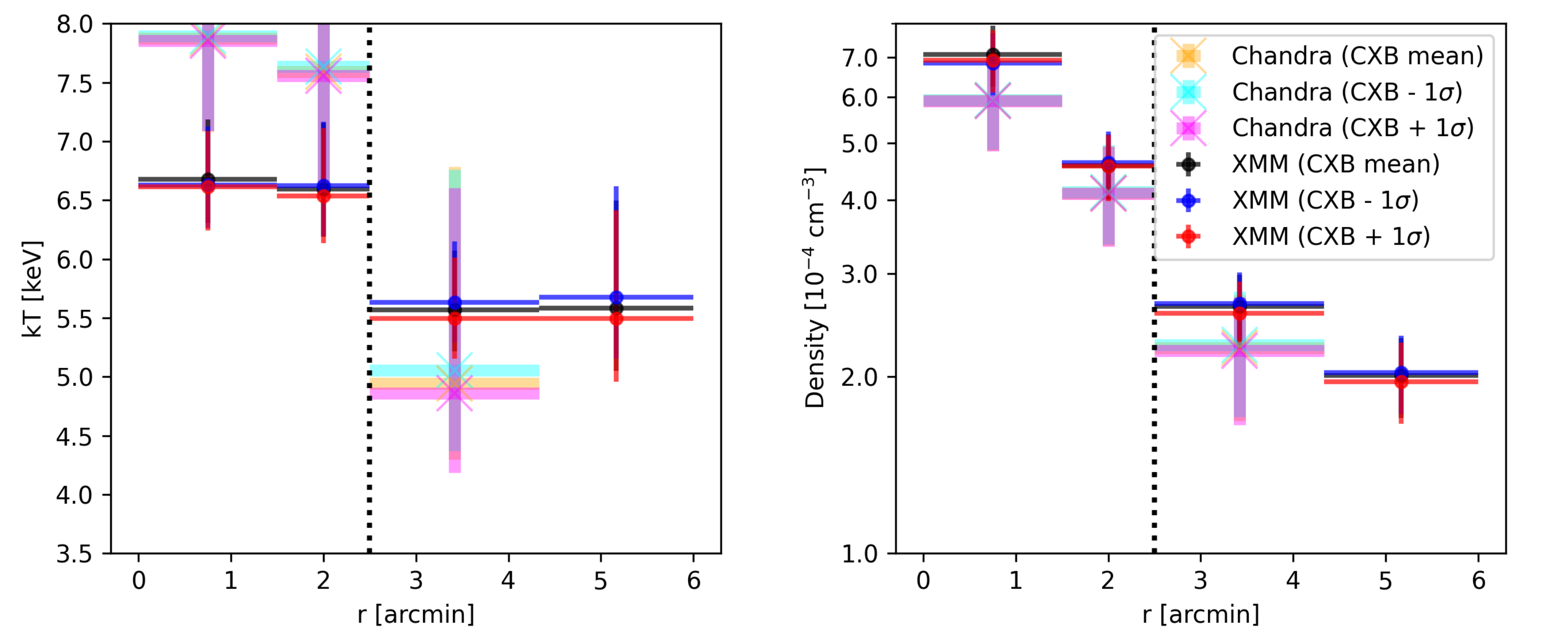}
\caption{ The kT spatial distribution (left) and density spatial distribution (right) were obtained by fitting spectra extracted from the magenta sectors from \textsl{XMM-Newton} and \textsl{Chandra}. The black dashed line marks the location of the shock front, approximately 2.3$'$ from the X-ray brightness peak of 1E2215, where a temperature and density jump can be observed. The fitted values are represented by black, red, and blue points with error bars (yellow, cyan, magenta points for \textsl{Chandra}), corresponding to medium, low, and high CXB fluxes, respectively. The high and low CXB fluxes were achieved by adding or subtracting the 1$\sigma$ statistical CXB flux error, as recorded in Tab.\ref{tab:cxb-paras}. Variations in the fitted kT and densities due to CXB uncertainties remained within the 1$\sigma$ statistical error, indicating that the temperature and brightness jumps are robust against these background fluctuations.
\label{fig:Tprof}}
\end{figure}

To determine the $1\sigma$ statistical uncertainties on the fitted parameters, we used the Monte Carlo chain command \texttt{chain} with a total chain length of 50,000 steps, discarding the first 5,000 steps to avoid early fluctuations. The resulting kT and density values are provided in Tab.\ref{tab:spec_props} and Tab.\ref{tab:spec_props_chandra}, and the kT profile is shown in Fig.\ref{fig:Tprof}. The kT profile reveals a noticeable jump at the radius marked by the black dashed vertical line around $2.3'$, which coincides with the surface brightness edge. This strongly supports our high-confidence conclusion that the observed density discontinuity in the surface brightness profile corresponds to a shock front, with a temperature jump ratio of \(T_{\rm in}/T_{\rm out} = 1.22 ^{+0.13}_{-0.14}\) and \(T_{\rm in}/T_{\rm out} = 1.55 ^{+0.37}_{-0.38}\) for \textsl{XMM-Newton} and \textsl{Chandra} respectively. The temperature jump remains consistent within the 1$\sigma$ statistical error across different CXB fluxes.

\section{Discussion}
\subsection{Mach number}

As shown in Fig.\ref{fig:sbp-prof}, a surface brightness discontinuity is observed at a radius of approximately 2.3$'$, corresponding to about 251.5\,kpc from the X-ray surface brightness peak of 1E2215. Assuming adiabatic compression, the Mach number of the brightness jump can be calculated using the Rankine-Hugoniot jump condition \citep{Landau59}. By plugging the surface brightness jump ratio $r$ into Eqn.\ref{eqn:sbp-mach}, we calculate the Mach number to be \( \mathcal{M} = 1.22 \pm 0.05 \) for \textsl{XMM-Newton}
, where \( \gamma = 5/3 \) is used for a monoatomic gas.

\begin{equation}
\label{eqn:sbp-mach}
    \mathcal{M} = \left[ \frac{2r}{\gamma + 1 - r \left( \gamma - 1 \right)} \right]^{\frac{1}{2}}
\end{equation}

Additionally, the Mach number can also be estimated from the temperature jump derived from spectral fitting, as expressed by Eqn.\ref{eqn:spec-mach}, where $kT_{\rm pre}$ and $kT_{\rm post}$ represent the pre-shock and post-shock temperatures, which gives \( \mathcal{M} = 1.25^{+0.11}_{-0.17} \) and \( \mathcal{M} = 1.53^{+0.36}_{-0.34} \) for \textsl{XMM-Newton} and \textsl{Chandra} respectively. The Mach numbers derived from both the surface brightness jump and the temperature jump are in strong agreement, indicating that the observed temperature jump can be largely explained by adiabatic compression resulting from the shock. Using the shock Mach number and the sound speed calculated from the pre-shock temperature (\( c_s \sim 1220\ \rm km\ s^{-1} \)), we estimate the shock speed to be \( v_s \sim 1500\ \rm km\ s^{-1} \). Assuming the shock propagates with a uniform speed from the cluster center, we derive an estimated shock age of approximately 280\,Myr.

\begin{equation}
\label{eqn:spec-mach}
    \frac{T_{\rm post}}{T_{\rm pre}} = \frac{\left[ 2 \gamma \mathcal{M}^2 - \left( \gamma -1  \right) \right] \left[ \left( \gamma -1 \right)\mathcal{M}^2 +2 \right]}{\left( \gamma + 1 \right)^2 \mathcal{M}^2}
\end{equation}

We did not observe any significant radio features near the shock front, as indicated by the white contours in Fig.\ref{fig:xmm-zoomin}. This absence may be attributed to a few possible reasons: either the majority of the cooling cosmic ray particles responsible for producing synchrotron emission are not present in the southeast region, or the shock is too weak to efficiently heat the electrons. This is consistent with the observed shock Mach number, \( \mathcal{M} \sim 1.2 \), which is lower than the theoretical minimum Mach number of \( \mathcal{M} \sim 2.2 \) required for efficient particle acceleration \citep{Ha18, Jacco21, Hoeft11}.

\subsection{The possible origin of the shock}

The shock discovered in this study may represent the spatial extension of the equatorial shock previously identified in this system, as the speed and age of the shock found in this paper are similar to those of the equatorial shock. The propagation speed of the previously discovered equatorial shock is approximately \( \sim 1740\ \rm km\ s^{-1} \), with an age of about 250\,Myr \citep{Gu19}, while the shock discovered in this paper has a speed of \( \sim 1500\ \rm km\ s^{-1} \) and an age of around 280\,Myr. The opening angle of the equatorial shock front can be large (\citealt{Ricker2001,Akahori10,Paul2011}). Therefore, it is possible that the extension of the equatorial shock is observed at this direction.Moreover, the age and the distance of the shock front is consistent with the theory (see figure 5 in \citealt{Ha18}).



The shock may also result from an ongoing merger within 1E2215, which is a non-cool-core, dynamically unrelaxed system. This merger could, in principle, be independent of the major merger between 1E2215 and 1E2216 (such as in \citealt{Eckert14,Feretti12,Russell14}). However, current data is insufficient to confirm its presence.


\section{Conclusion}

In this paper, we present a study of the 1E2215-2216 galaxy cluster pair using new data from \textsl{XMM-Newton} observations. The extended exposure allowed us to identify a new shock located 2.3$'$ south of the surface brightness peak of 1E2215. The shock is confirmed by a surface brightness jump of \(1.33 \pm 0.07\) and a temperature jump of \(1.22^{+0.13}_{-0.14}\), with the Mach numbers estimated independently from the surface brightness and temperature jumps being \(1.22 \pm 0.05\) and \(1.25^{+0.11}_{-0.17}\) respectively in \textsl{XMM-Newton}, which are in agreement. The \textsl{XMM-Newton} results also in line with \textsl{Chandra} results. Given that the age and speed of this shock are similar to the previously discovered equatorial shock, this new shock may represent a physical extension of the equatorial shock. Alternatively, it could be caused by the infall of a group, unrelated to the collision of these two galaxy clusters.

\begin{acknowledgements}
The authors thank the anonymous referee for their comments that have improved this paper. The authors wish to thank Jelle de Plaa, Zhenlin Zhu, Anwesh Majumder, and Jiejia Liu for helpful discussion, and the Tsinghua Astrophysics High-Performance Computing platform at Tsinghua University for providing computational and data storage resources. This work was supported in part by the National Natural Science Foundation of China through Grant 11821303, and by the Ministry of Science and Technology of China through Grant 2018YFA0404502. YLC wishes to acknowledge support from the China Scholarship Council. This paper employs a list of Chandra datasets, obtained by the Chandra X-ray Observatory, contained in the Chandra Data Collection ~\dataset[DOI: 10.25574/cdc.370]{https://doi.org/10.25574/cdc.370}.
\end{acknowledgements}

\begin{appendix} 

\section{Find best model for surface brightness profile}
\label{app:sbp-mdls}

We conducted a BIC test to find whether the broken power law or beta model is the best model of the surface brightness profile we extracted from the \textsl{XMM-Newton} and \textsl{Chandra} data. The results are shown in Tab.\ref{tab:bic}. According to the \citet{kass1995}, the model is better if the BIC value is smaller. The $\Delta \rm BIC$ value in Tab.\ref{tab:bic} is defined as $\Delta \rm BIC = BIC_{bkn}-BIC_{beta}$. The positive $\Delta \rm BIC$ means beta model better than broken power law, and $10 >| \Delta \rm BIC|>6$ means the evidence is strong, $|\Delta \rm BIC|>10$ means the evidence is definitive. According to the table, only sector of 230$^{\circ}$-280$^{\circ}$ shows strong evidence of surface brightness edge in \textsl{XMM-Newton} data.

\begin{deluxetable*}{ccccccc}
\label{tab:bic}
\tablecaption{ $\Delta $BIC between the broken power law and beta models.}    
\tablehead{\colhead{Instrument} & \colhead{20-70} & \colhead{70-120} & \colhead{170-230} & \colhead{230-280} & \colhead{280-330} & \colhead{330-20}} 
    
\startdata
\textsl{Chandra} &  
6.41 &
10.41 & 
-0.93 &
1.64 &
12.29 &
12.81 \\
\textsl{XMM-Newton} & 
70.87 & 
6.88 & 
24.70 & 
-6.23 &
55.64 &
86.01 \\
\enddata 
\end{deluxetable*}

\section{Background model settings for spectral analysis}
\label{app:spec-mdls}

\begin{figure}[ht!]
\centering
\includegraphics[width=1\textwidth]{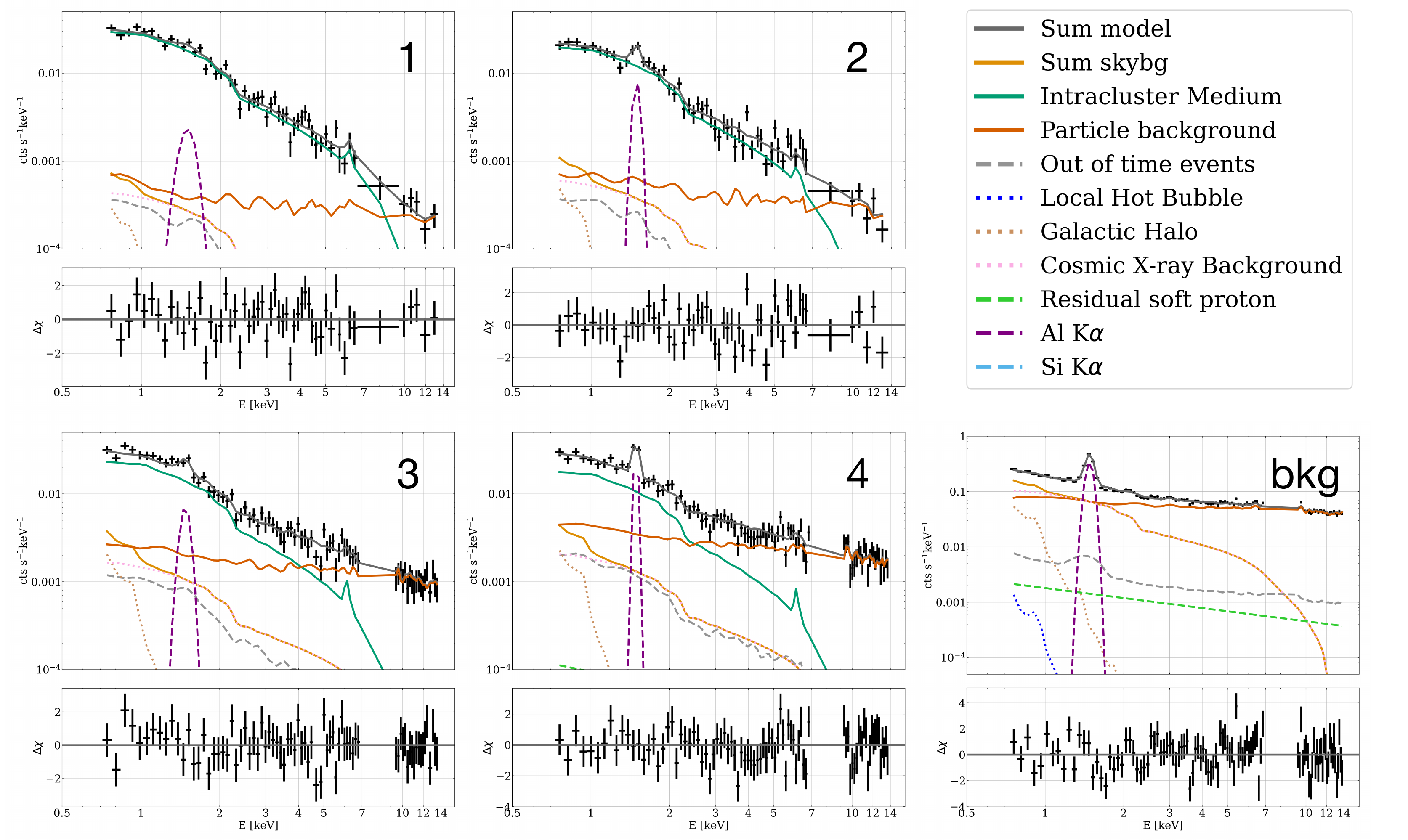} 
\caption{The fitted spectra for the source and sky background regions of \textsl{XMM-Newton} for CXB fixed at middle value. The data points are shown in black with corresponding error bars. The fitted models consist of multiple components, each represented by a different color curve. 
The quiescent soft proton contribution is two orders of magnitude lower than the data, making it invisible in the plots. 
\label{fig:spec-src}}
\end{figure}

\begin{figure}[ht!]
\centering
\includegraphics[width=1\textwidth]{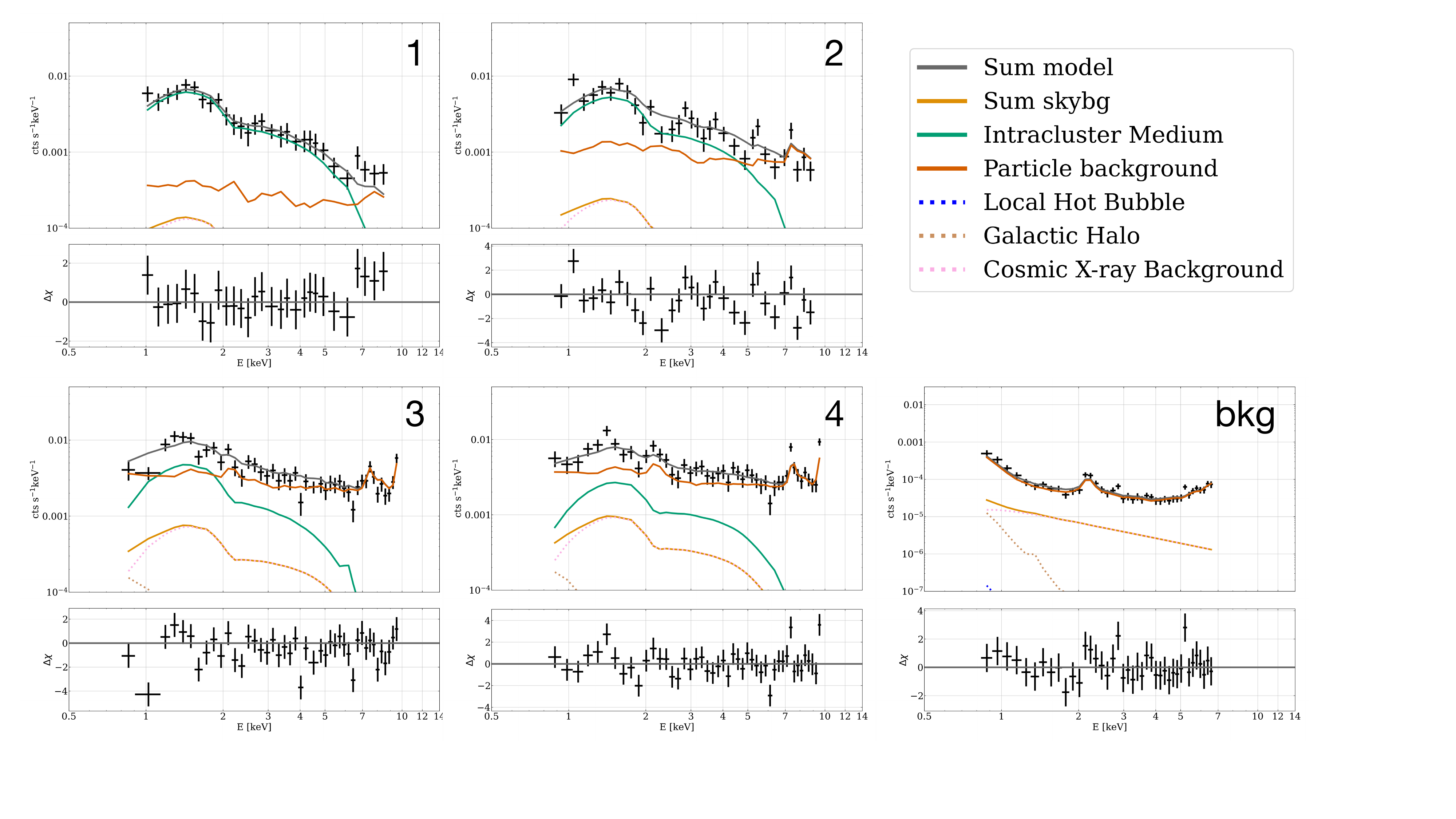} 
\caption{The fitted spectra for the source and sky background regions of \textsl{Chandra} observation 20778. The line labels are the same as in Fig.\ref{fig:spec-src}.
\label{fig:spec-chandra}}
\end{figure}

The background model used in this analysis includes several components: Galactic Halo (GH), Local Hot Bubble (LHB), Cosmic X-ray Background (CXB), and Non-X-ray Background (NXB), with Galactic absorption taken into account. To ensure proper modeling, the areas of different regions are considered by rescaling the corresponding model spectra. Example figures illustrating the fitted spectral components for both source regions and sky background regions are shown in Fig.\ref{fig:spec-src} and Fig.\ref{fig:spec-chandra} for \textsl{XMM-Newton} and \textsl{Chandra}. The components used to derive the spectral models are outlined below.

\begin{itemize}

\item{\textit{The galactic absorption:}} The Galactic hydrogen column density used in this study was obtained from the Swift survey \citep{willingale13}, which takes into account both atomic and molecular hydrogen. For the 1E2215/2216 region, the weighted total hydrogen column density is \( N_{\rm H_{\rm tot}} = 7.23 \times 10^{20}\ \rm cm^{-2} \). This value was used for spectral modelling to correct for Galactic absorption along the line of sight.

\item{\textit{The foreground components:}} We extracted the spectra from the sky background region of the \textsl{XMM-Newton} data, which is located outside the $2r_{200c}$ of the two galaxy clusters, as shown by the pink solid sector in Fig.\ref{fig:xmm-overall}. Additionally, we extracted the diffuse background spectra from the ROSAT All Sky Survey (RASS) data, using an annular region with an inner radius of 18$'$ and an outer radius of 30$'$, which is also located outside the $2r_{200c}$ of the clusters. To constrain the properties of the Galactic Halo (GH) and the Local Hot Bubble (LHB), we fit the sky background spectra from both \textsl{XMM-Newton} and ROSAT in parallel with the source spectra. The Galactic Halo (GH) was modeled using an absorbed thin APEC plasma model, while the Local Hot Bubble (LHB) was modeled with an unabsorbed thin APEC plasma model. The metallicity of both the LHB and GH was fixed at $1Z_{\odot}$. For the LHB, the temperature was fixed at 0.11\,keV, and the normalization was allowed to vary \citep{snowden97}; for the GH, the preliminary temperature was set at 0.2\,keV \citep{Yoshino09}, both the temperature and normalization were allowed to vary during the fitting process. We defined the region out of $1.1r_{200c}$ in \textsl{Chandra} data as the background region, as shown in green circles in right panel of Fig.\ref{fig:xmm-zoomin}. The fitted foreground parameters in \textsl{Chandra} are consistent with \textsl{XMM-Newton}. The best-fit parameters of foreground are recorded in Tab.\ref{tab:spec_bkgpara}.

\begin{figure}[ht!]
\plotone{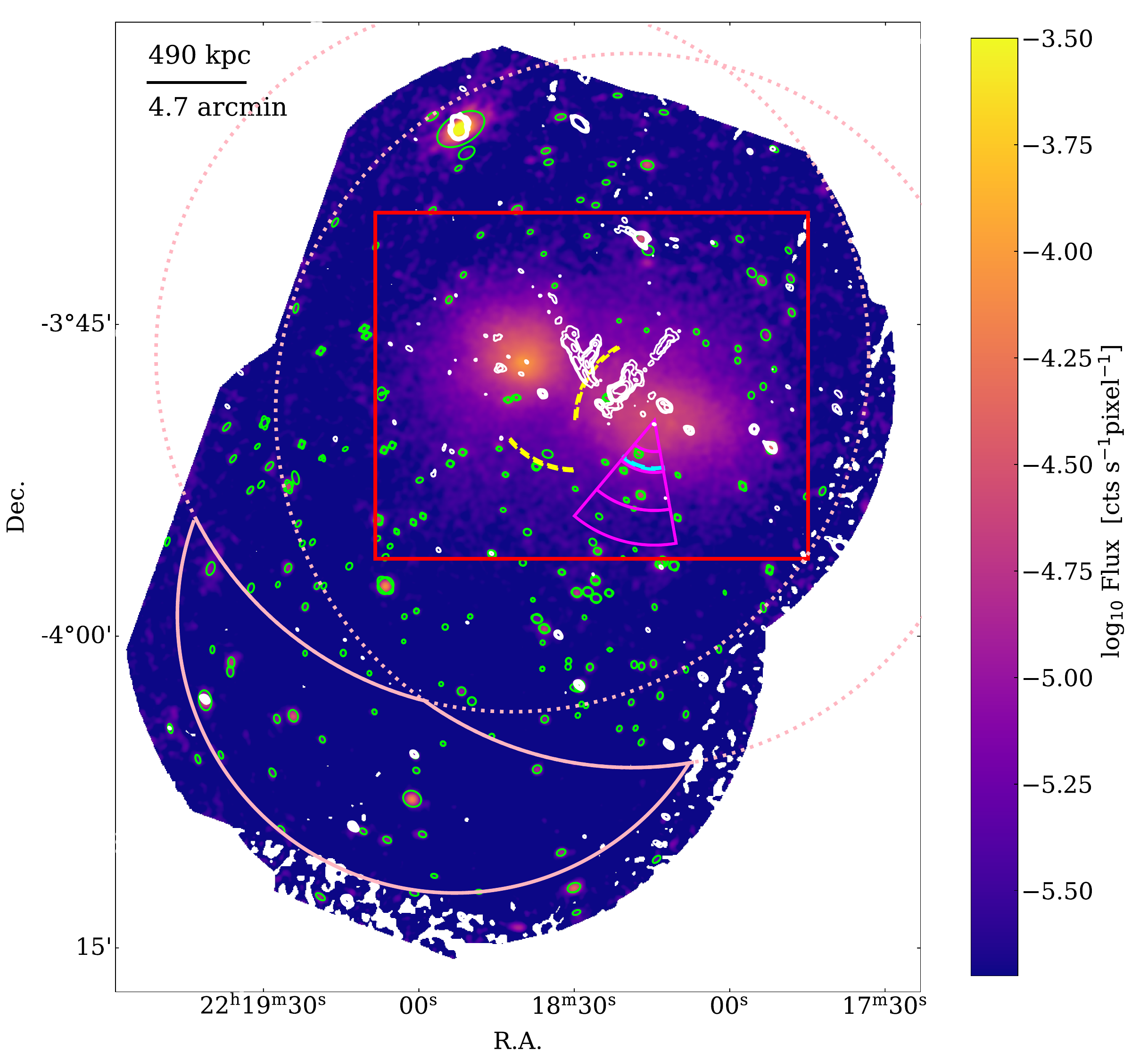}
\caption{ The [0.5-2.0]\,keV \textsl{XMM-Newton} image of the entire dataset is shown, with particle background subtracted and vignetting effects corrected. The magenta sectors indicate the regions used for spectral fitting in \textsl{XMM-Newton} data, while the pink solid sector marks the sky background region. The pink dotted circles are the $2r_{200c}$ of the two clusters. The green ellipses represent the excluded point source regions. The red box highlights the zoom-in area, which is shown in Fig.\ref{fig:xmm-zoomin}. The image has been slightly smoothed using a Gaussian kernel with a full width at half maximum (FWHM) of 1.5 pixels, i.e, 3.8$''$ or 2.3\,kpc.
\label{fig:xmm-overall}}
\end{figure}

\begin{deluxetable*}{cccccc}
\label{tab:spec_bkgpara}  
\tablecaption{Best fitted parameters of the LHB and GH }    
\tablehead{\colhead{Component} & \colhead{Model} & \colhead{Parameters} & \colhead{Values} & \colhead{Unit} & \colhead{Status}}
\startdata
LHB & APEC & T & 0.11 &\,keV & Fixed \\
    &      & norm & $(4.37 \pm 0.10) \times10 ^{-7}$ & arcmin$^{-2}$ & Freed \\
GH  & TBabs $\times$ APEC & T & $0.169 \pm 0.003$  &\,keV & Freed \\
    &      & norm & $(2.27 \pm 0.12) \times 10^{-6}$ &  arcmin$^{-2}$ & Freed \\
\enddata 
\tablecomments{The normalization here is defined as $\frac{10^{-14}}{4\pi\left[D_{A}\left(1+z\right)\right]^2\int n_e n_H dV}$, where $D_A$ is the angular diameter distance to the source (cm), $dV$ is the volume element ($\rm cm^3$), and $n_e$, $n_H$ are the normalized electron and H densities ($\rm cm^{-3}$). The density normalizations are normalized to $1\ \rm arcmin^2$. }
\end{deluxetable*}

\begin{deluxetable*}{ccccc}
\label{tab:cxb-paras}  
\tablecaption{CXB fluxes and uncertainties}    
\tablehead{\colhead{1} & \colhead{2} & \colhead{3} & \colhead{4}}
\startdata
$4.85 \pm 3.81$ & $4.68 \pm 3.09$ & $4.42 \pm 1.91$  & $4.43 \pm 1.19$ \\ 
\enddata 
\tablecomments{The sectors from inner to outer radii are marked as 1,2,3,4. The unit of the CXB flux is $10^{-15}\ \rm erg\ s^{-1}\ cm^{-2}\ arcmin^{-2}$ in [0.5-7.0]\,keV. }
\end{deluxetable*}

\item{\textit{The non X-ray background:}} The non-X-ray background (NXB) spectra were smoothed using the \texttt{bkgsmooth} function in the \texttt{pyspextools} code \citep{jdeplaa2023}, which applies the Savitzky–Golay filter. For the MOS data, the window length was set to 500 with a polynomial degree of 1, while for the PN data, a window length of 140 and a polynomial degree of 3 were used, as the NXB spectra for the PN detector are more complex and contain more spectral lines. For \textsl{Chandra} data, the window length and the polynomial degree are set as 11 and 3 respectively. The smoothed instrumental background spectrum was then loaded as a table model in XSPEC using the \texttt{atable} task. To account for background uncertainties, we scaled the overall background model by comparing the ratio of the background count rate to the data count rate in the [10-14]\,keV band, and allowed the background normalization to vary by 10\% to account for potential uncertainties in the background estimation.

\item{\textit{The soft proton and OOT spectra of \textsl{XMM-Newton} data:}} The quiescent soft proton background was fitted using a broken power law, folded with a dummy response because soft proton particles. The initial index of the broken power law was set to 0.6, with variation allowed between 0.1 and 2.5. The break energy was fixed at 3\,keV but allowed to vary. 
To address prominent instrumental lines, such as the Al and Si lines in MOS, and the Al line in PN, which appear in the [1.2–1.7]\,keV band, Gaussian models were used to fit the lines, with the energy of the Al line fixed at 1.49\,keV, the Si line at 1.75\,keV, and their widths were allowed to vary within a range of 0.2\,keV.
The OOT spectra in PN were smoothed using the Savitzky–Golay filter, with a window length of 140 and a polynomial degree of 3. The smoothed OOT spectra were then scaled by a factor of 0.023, as the PN detector was operated in Extended full-frame mode during the observations. All components, as described above, were loaded using dummy responses provided by the ESAS software, and the effective area files were not included in the fitting process.

\item{\textit{The Cosmic X-ray background:}} In \textsl{XMM-Newton}, due to the spatial variation in detector sensitivity caused by vignetting effects and the background cluster emission, sources close to the center but out of cluster regions are effectively identified. Applying a uniform detection limit based on the detector edge would fail to exclude these resolved point sources, artificially inflating the measured CXB that could be removed.
To address this issue, we generated a sensitivity map \citep{Huang24}, which allows for pixel-dependent detection limits (see Fig.\ref{fig:sns-logNS} on the right). This method enables the detection threshold to be appropriately adjusted across the detector, helping to effectively reduce uncertainties in the resolved CXB fluxes.

\begin{figure}[ht!]
\plotone{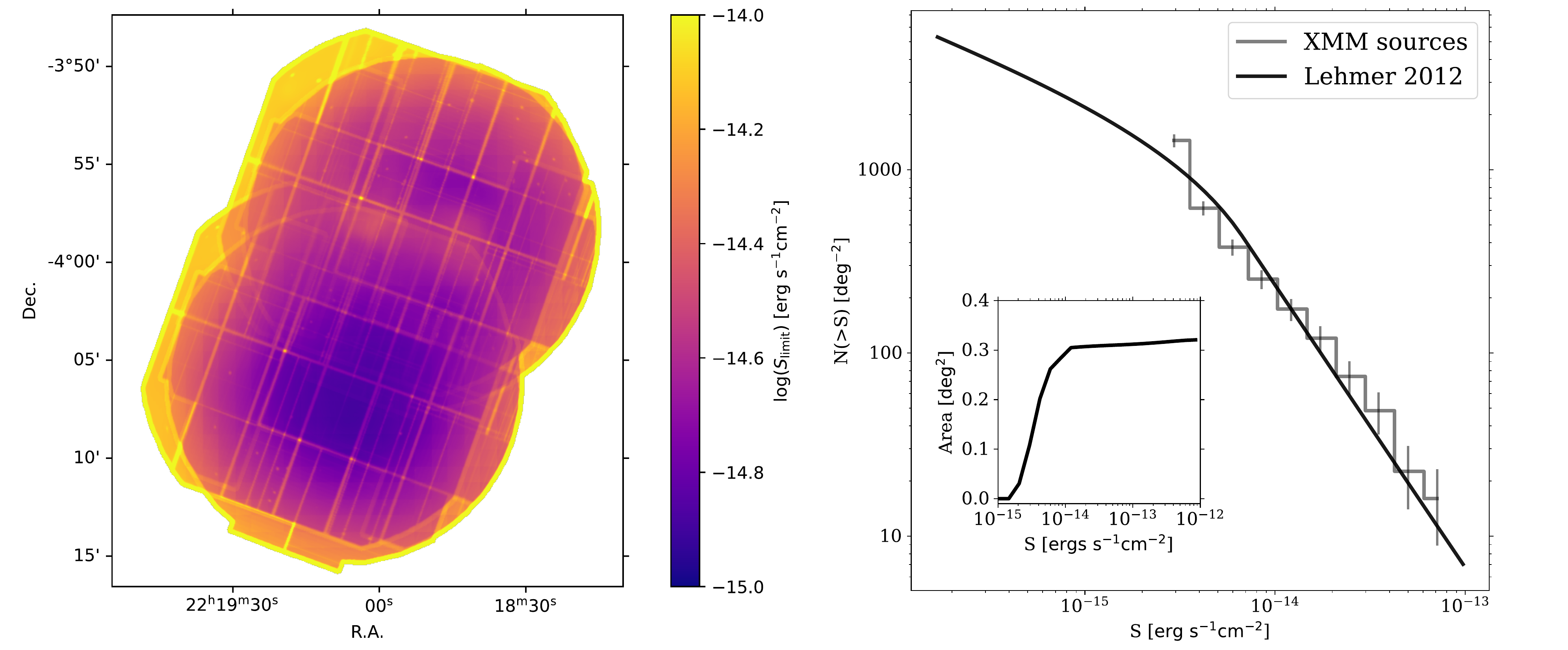}
\caption{Left: The \textsl{XMM-Newton} sensitivity map of 1E2215/2216. Right: The logN-logS curve in the \textsl{XMM-Newton} FoV. In the left panel, the sensitivity $S_{\rm limit}$ improves from the edge to the center of the FoV, but deteriorates in the cluster region. In the right panel, the black stairs show the number of point sources in \textsl{XMM-Newton} with fluxes higher than a certain value on the x-axis. The black curve represents the logN-logS relation in the [2.0-8.0]\,keV band of CDF-S. The inset in the right panel shows the effective area versus the sensitivity flux, which is the denominator of the AGN number in each flux bin to correct for completeness. 
\label{fig:sns-logNS}}
\end{figure}

For each pixel, we integrated the LogN-LogS relation from zero to the detection limit. We adopted the LogN-LogS relation from \citealt{Lehmer12}, which aligns well with our data (see right panel of Fig.\ref{fig:sns-logNS}), using cxbtools \citep{cxbtool2017}. Since we are primarily interested in the CXB flux contribution from background AGNs, we focused on the LogN-LogS curve in [2.0-8.0]\,keV band. We then generated the sensitivity map for the [2.0-4.5]\,keV band to be consistent with the fluxes recorded in the 4XMM-DR13 catalog. To convert the source fluxes in the [2.0-4.5]\,keV band to the [2.0-8.0]\,keV band, we assumed a counts-to-flux factor based on an absorbed power law model with a photon index of 1.41. This photon index is considered to represent the mean spectral energy distribution (SED) of the X-ray background, as outlined by \citep{moretti09}.

As shown in Fig.\ref{fig:sns-logNS} (right), we calculated the number of sources with fluxes above a certain threshold and divided it by the area in the sensitivity map (shown in inset) corresponding to fluxes greater than that threshold to correct for completeness (black stairs). By comparing the LogN-LogS relation in the 1E2215/2216 field observed by \textsl{XMM-Newton} with the LogN-LogS curve from the CDF-S (Chandra Deep Field Survey) \citep{Lehmer12}, we found that the curve in the 1E2215/2216 field (black stairs) aligns with the LogN-LogS curve of CDF-S (black curve). Therefore, we directly integrated the LogN-LogS curve of CDF-S from zero to the detection limit without scaling to calculate the unresolved Cosmic X-ray Background (CXB) flux. In the spectral analysis, we fitted the CXB using an absorbed power-law model with a photon index of 1.41. For every pixel in the \textsl{XMM-Newton} data, we integrated the LogN-LogS curve and recorded the CXB flux values for different regions in Tab.\ref{tab:cxb-paras}. The CXB flux for the sky background is $12.63 \pm 0.74 \times 10^{-15}\ \rm erg\ s^{-1}\ cm^{-2}\ arcmin^{-2}$. We fitted the spectra three times, setting the CXB flux to the middle flux value, the CXB - $1\sigma$ flux value, and the CXB + $1\sigma$ flux value, and fixed the CXB flux during the fitting process. 

In \textsl{Chandra} data, we expected the effects on results caused by CXB systematic uncertainties is small, therefore we take the CXB flux and its statistical errors constrained by sky background to fit the source spectra, which is $4.07 \pm 0.76 \times 10^{-15} \rm\ erg\ s^{-1}\ cm^{-2}$ in [0.5-7.0]\,keV.
 
\end{itemize}
\end{appendix}

\bibliography{reference.bib}{}
\bibliographystyle{aasjournal}
\end{document}